\documentclass{aa}

\usepackage{graphicx}
\usepackage[varg]{txfonts}
\usepackage{comment}

\usepackage{amsmath}

\newcommand{\vect}[1]{{\mathbf{{#1}}}}

\newcommand{\rsun}{\mathrm{r}_{\odot}}

\newcommand{\kappaang}{\kappa_{\phi}}

\title{Solar Energetic Particle event onsets at different heliolongitudes: The effect of turbulence in Parker Spiral Geometry}
\titlerunning{SEP onsets in Parker spiral turbulence}
\authorrunning{Laitinen et al} 

\author{T.~Laitinen \and S.~Dalla \and C.~O.~G.~Waterfall \and A.~Hutchinson} 
\institute{Jeremiah Horrocks Institute, University of Central Lancashire, UK}

\date{Received date /
Accepted date }

\abstract{Solar energetic particles (SEPs), accelerated during solar eruptions, are observed to rapidly reach a wide heliolongitudinal range in the interplanetary space. To access these locations, the particles must have either been accelerated at a wide particle source, or propagated across the mean Parker spiral magnetic field.}%
    {We study the propagation of SEPs in a new model of heliospheric turbulence which takes the spiral geometry of the average magnetic field into account, to evaluate how this improved description affects the SEP path lengths and the overall evolution of SEP intensities at 1~au.}%
    {We use full-orbit test particle simulations of 100-MeV protons in a novel analytic model for turbulent magnetic field in the Parker spiral geometry, where the turbulence is dominated by modes that are transverse and 2-dimensional with respect to the Parker spiral direction.}%
    {We find that the particles spread along the meandering field lines to arrive at a 60$^\circ$ heliolongitudinal range at 1~au heliocentric distance within an hour of their injection at the Sun, consistent with the heliolongitudinal extent of the meandering field lines. The SEP onset times are asymmetric with respect to the location connected to the source along the Parker spiral, with westward locations seeing earlier arrival and higher peak intensity. The inferred path length of the first-arriving particles is 1.5-1.7~au, 30-50\% longer than the Parker spiral, and 20\% longer than the length of the random-walking field lines. Subsequently, the SEP distribution broadens, consistent with diffusive spreading of SEPs across the field lines.}{Our results indicate that due to the nature of interplanetary turbulence, SEPs can propagate rapidly across the mean Parker Spiral field to arrive at wide range of longitudes, even without a wide particle source. The modelled SEP onset times, the peak intensity and subsequent heliolongitudinal evolution replicate several observed SEP event features. Further studies are be required to investigate the relative importance of interplanetary transport and source size in different turbulence environments.}

\begin{document}

\maketitle


\section{Introduction} \label{sec:intro}

Solar energetic particles (SEPs) are accelerated during solar eruptions.
The acceleration processes of SEPs are typically divided to flare-related processes and those related to the shock wave driven by coronal mass ejections (CMEs) in the corona and interplanetary space. The traditional, so-called impulsive/gradual classification of SEP events considers flare-related events as small, short-lived, rich in heavy elements, and narrow in heliolongitudinal extent, whereas the gradual, proton rich events are seen at wide range of heliolongitudes, and can last for up to over a week \citep[e.g.][]{Reames1999}. This classification has been challenged particularly by the recent multi-spacecraft observations of heavy element-rich SEP events that have a wide heliolongitudinal extent \citep{Wiedenbeck2013, Cohen2014, Cohen2017}, similar to that of typical gradual SEP event heliolongitudinal extents \citep[e.g.][]{Lario2006, Lario2013, Richardson2014}. Several ideas have been presented to explain the observation of heavy elements over broad regions in heliolongitudes, such as coronal spreading of magnetic field lines \citep[e.g.][]{Liewer2004_coronalopening}, sympathetic flaring \citep[e.g.][]{Schrijver2011_sympathetic}, reacceleration of remnant flare particles by CME-driven shock waves \citep[e.g.][]{Mason1999, Reames1999, Desai2003}, and cross-field propagation of SEPs in the interplanetary space due to solar wind turbulence.

The effect of turbulence on SEP propagation in the interplanetary space is typically described as pitch angle diffusion along and spatial diffusion across the mean Parker spiral direction \citep[e.g.][]{Parker1958,Jokipii1966}. Earlier modelling of SEP observations concentrated on SEP propagation along the mean Parker spiral \citep[e.g.][]{Kallenrode1993, Torsti1996, Droge2000ApJ,AgEa09} as, based on the so-called Palmer consensus \citep{Palmer1982}, the cross-field effects due to turbulence were considered negligible. 

Recently, the significance of stochastic cross-field propagation of SEPs has gained more attention. Most models have concentrated on studying the effect of the turbulence as diffusion across the mean field direction on SEP intensities \citep[e.g.][]{Zhang2009, Droge2010, He2011, Strauss2015}. However, \citet{LaEa2013b} noted that in the timescales of SEP propagation to Earth, their early transport is dominated by deterministic propagation along stochastically meandering field lines, rather than stochastic motion relative to the mean magnetic field. \citet{LaEa2016parkermeand} implemented this new description of transport into a Fokker-Planck framework combining field line and particle diffusion, and showed that the combination of the initial deterministic propagation and time-asymptotic cross field diffusion processes resulted in a particle population that extended rapidly to wide range of longitudes. Subsequently, the modelled SEP event continued to broaden in heliolongitude at slower pace, depending on turbulence strength \citep{LaEa2018}.

The diffusion approach for particle or field line cross-field diffusion, however, is not able to address an important element of the SEP transport: the length of the stochastically meandering field lines, crucial for connecting the solar SEP source to the analysis of the observed SEP onsets at the observing spacecraft. Long apparent path lengths of SEPs obtained from velocity dispersion analysis (VDA) have been reported in several studies \citep[e.g.][]{Paassilta2017, Paassilta2018, LeskeEa2020}. Recent simulation studies have shown that the turbulence in general lengthens the mean length of interplanetary field lines \citep[e.g.][]{Pei2006,MoradiLi2019,Chhiber2020Pathlengths}. In \citet{LaDa2019pathlength} we used a random-walk model to demonstrate that the lengthening of the field lines is asymmetric in heliolongitude, with longer field line lengths connecting an observer located east of the Parker spiral magnetic field connected to the field line source, as compared to a path connecting a westward observer. 

In a more recent work, we introduced a novel Parker spiral geometry turbulence model, where for the first time the dominant 2D turbulence component wave modes are 2-dimensional with respect to the Parker spiral background magnetic field \citep{LaEa2022_ParkerTurb}. Analysis of field-line lengths using this model agreed with the asymmetric distribution of field line lengths found by the random-walk model of \citet{LaDa2019pathlength}.

Our earlier work only investigated field line lengths, and did not include the effect of the turbulent magnetic field fluctuations on SEP propagation along the meandering field lines. In this paper, we use full-orbit test particle simulations to trace SEP propagation in Parker spiral turbulence model to investigate how turbulent fluctuations affect the arrival of the first particles to different longitudes at 1~au. From the simulations we derive the longitudinal dependence of onset and peak times of SEP intensities for comparison with observations \citep[e.g.][]{Richardson2014}. We describe our model parameters and assumptions in Section~\ref{sec:models}, and present our results in Section~\ref{sec:results}, and discuss their implications and draw our conclusions in Section~\ref{sec:discussion}.

\section{Model} \label{sec:models}

In this work, we integrate the full equation of motion of energetic particles from the source region near the Sun, and record the particles as they pass through a 1-au heliocentric sphere. We modified our Parker Spiral full-orbit 3D test particle code \citep{Dalla2005,Marsh2013} to include, in addition to the Parker spiral field, a turbulent magnetic field component. The turbulence model, fully described in \citet{LaEa2022_ParkerTurb}, is dominated by a transverse 2D component where the wavenumber vector and the magnetic field vector are normal to each other, and normal to the background Parker spiral field. In addition to the 2D component, also a slab-like component is included.  It should be emphasised that within the new version of the code, we do not include ad-hoc pitch angle scattering as in the previous versions \citep[e.g.][]{Marsh2013}: any changes of the particle velocity vector and position are solely due to the modelled magnetic field.

In this paper, we use the same parametrisation of the solar wind and turbulence as in \citet{LaEa2022_ParkerTurb}. The Parker spiral magnetic field is parametrised by the magnetic field at $1\;\rsun$ of ~1.78 gauss, the solar wind velocity $v_{sw}=400\;\mathrm{km/s}$ and solar angular rotation rate $\Omega=2.86533\times 10^{-6}\;\mathrm{rad}\; \mathrm{s}^{-1}$. The 2D and slab components are partitioned to have 80\%:20\% energy division. The spectra of both components have a flat energy-containing range between the largest scale $k_0=1/r$ and the breakpoint scales $\lambda_{2D}$ and $\lambda_{slab}$, where $r$ is the heliocentric distance, and Kolmogorov spectrum above the breakpoint scales. As in \citet{LaEa2022_ParkerTurb}, we use  $\lambda_{\perp}=0.04 (r/r_\odot)^{0.8} r_\odot$,  $\lambda_{slab}=2\lambda_{2D}$, and $\delta B^2=0.03 B_{p\odot}^2 (r/r_\odot)^{-3.3}$, where $B_{p\odot}$ is the magnitude of the Parker spiral magnetic field at $1\;\rsun$.

It should be noted that in this paper we include neither the convective electric field $\vect{E}=-\vect{v}_{sw}\times\vect{B}$ nor an electric field related to the turbulent fluctuations. For this reason, the particles retain their original energies, and do not experience the $\vect{E}\times\vect{B}$-drift which gives rise to the corotation drift \citep[e.g.][]{Dalla2013}. The effects of a convective and turbulent electric fields will be investigated in future work.

\section{Results}\label{sec:results}

\begin{figure}
  \resizebox{\hsize}{!}{\includegraphics{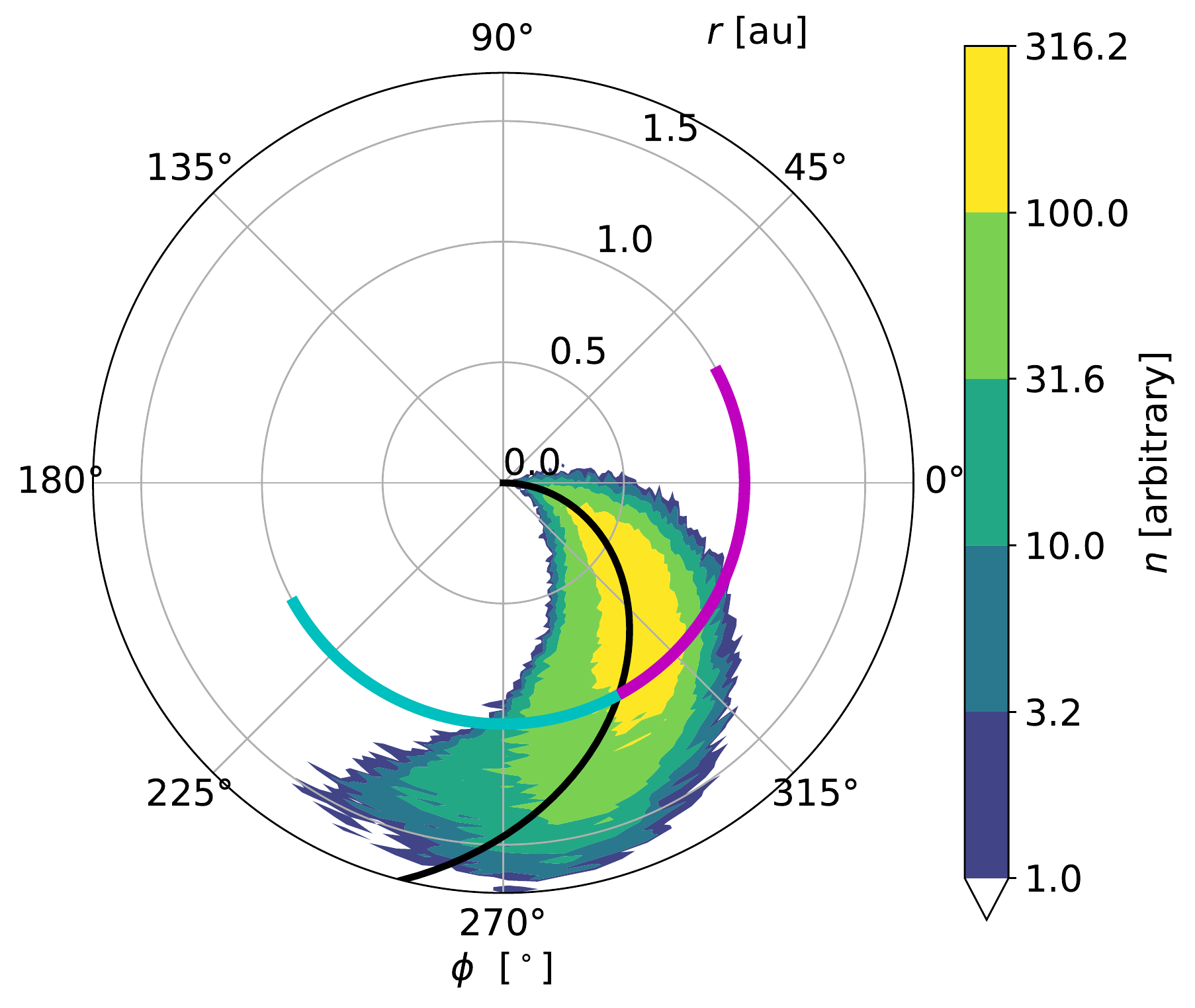}}
  \caption{Distribution of 100 MeV protons 1 hour after injection from a $8^\circ\times 8^\circ$ source at the solar equator at $\phi=0^\circ$, integrated over latitude and binned in longitude and radius bins. The black curve shows the nominal Parker spiral originating from the centre of the source region. The magenta and cyan arcs at $r=1$~au depict ranges for $\Delta\phi<0^\circ$ and and $\Delta\phi>0^\circ$, respectively (see Equation~(\ref{eq:dphi})).}\label{fig:rphi1hour}
\end{figure}

We simulate an impulsive injection of 100,000 100-MeV protons from a source of $8^\circ\times8^\circ$ heliolongitudinal and -latitudinal area centred at the solar equator and heliolongitude $\phi_{src}=0^\circ$ at 2~$\rsun$ heliocentric distance. The protons are traced for 48 hours as they propagate in the turbulent heliospheric magnetic field. In Figure~\ref{fig:rphi1hour}, we show the density of particles, integrated over latitude and binned over heliocentric distance and heliolongitude, after one hour of propagation. As we can see, the protons have spread from their original narrow $8^\circ$ source to over 60$^\circ$ heliolongitudinal range at 1~au, as a result of meandering magnetic field lines.

\begin{figure*}
\centering
\includegraphics[width=0.46\textwidth]{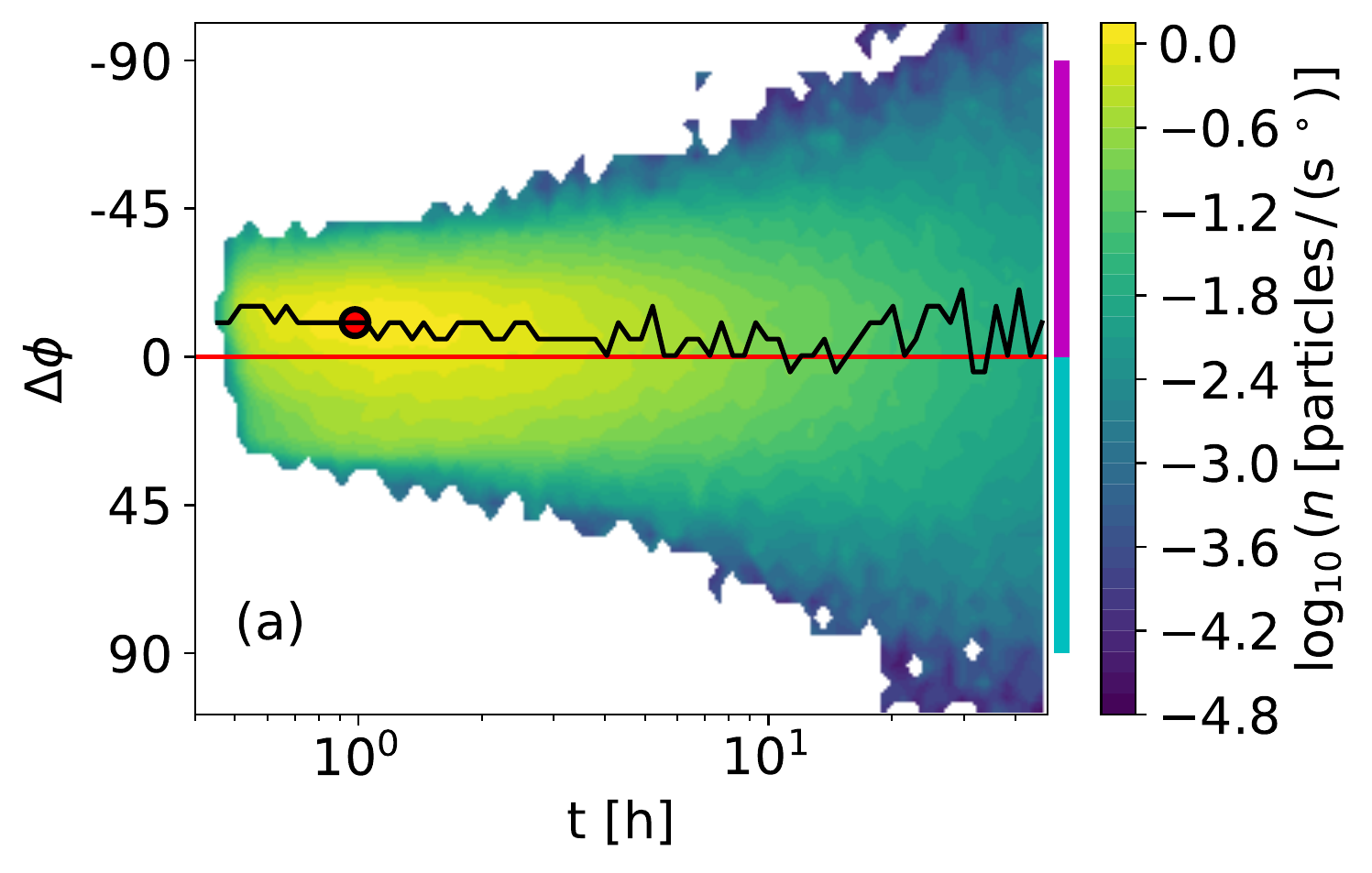}
\includegraphics[width=0.46\textwidth]{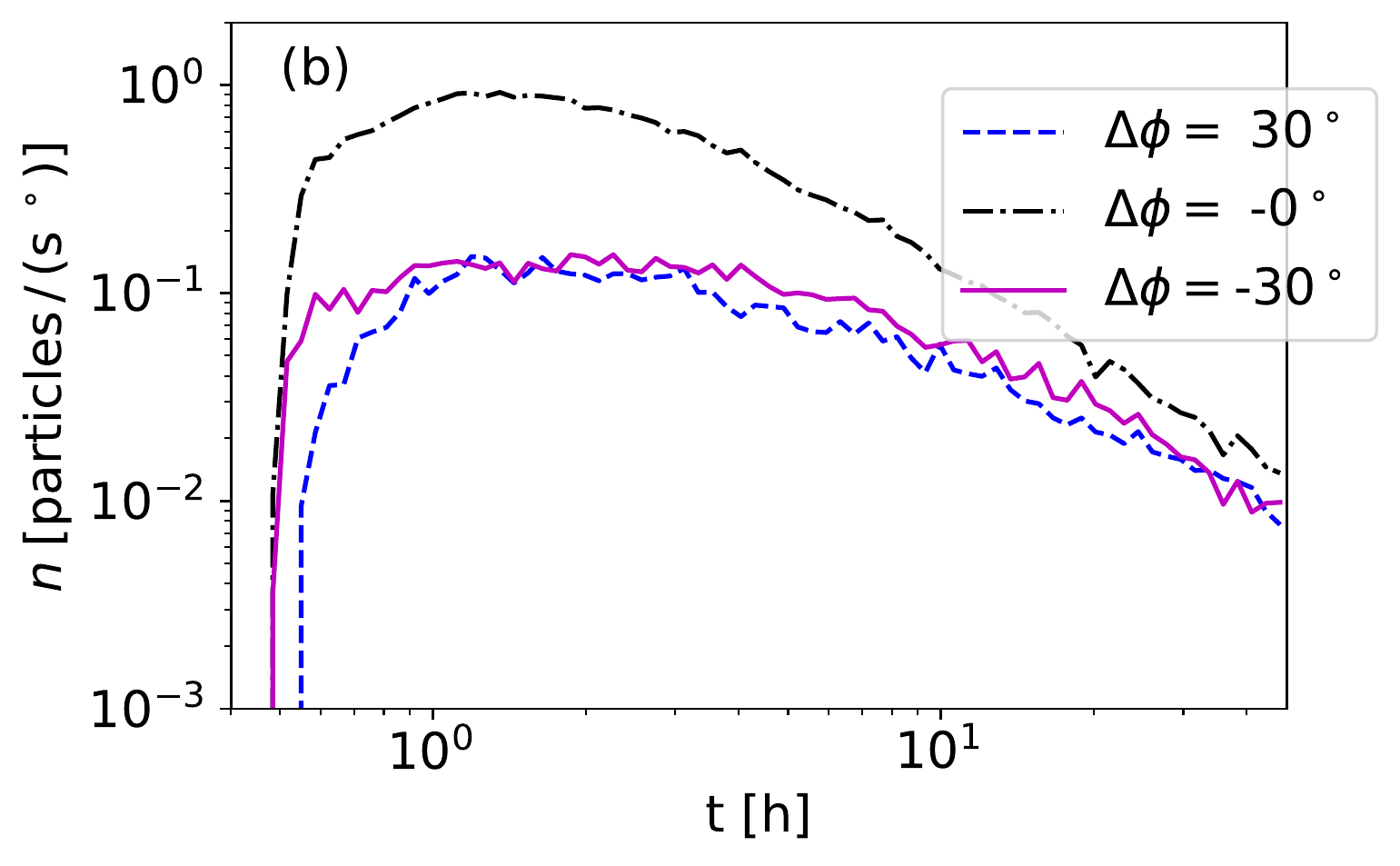}
\caption{(a) Contour of 100-MeV proton intensity evolution as a function of time and  $\Delta\phi$ at 1~au. The horizontal red curve depicts the longitude of the footpoint centred at the particle source at 2$\rsun$, $\phi_{src}$, the black curve the longitude of the peak intensity as a function of time, and the red circle the time and heliolongitude of the maximum intensity. The magenta and cyan ranges on the right correspond to those in Figure~\ref{fig:rphi1hour}. (b) Time evolution of the densities at three heliolongitudes.
\label{fig:phitimeint}}
\end{figure*}

In Figure~\ref{fig:phitimeint}, we investigate the temporal evolution of the heliolongitudinal particle density at 1~au. In panel (a), we show a contour plot of the density, as a function of time and relative heliolongitude
\begin{equation}\label{eq:dphi}
\Delta\phi=\phi_{src}-\phi_{fpt},    
\end{equation}
where $\phi_{fpt}$ is the footpoint longitude that connects an observer at $(r, \phi)$ to the solar surface along the Parker spiral magnetic field. For $v_{sw}=400$~km/s, the longitude at $r=1$~au for which $\Delta\phi=0^\circ$ is $\phi_{1au}=-61^\circ=299^\circ$: that is, an observer at that longitude is connected to the source $\phi_{src}=0^\circ$, along the black curve in Figure~\ref{fig:rphi1hour}. Observers located westwards from $\Delta\phi=0^\circ$ (magenta arc in Figure~\ref{fig:rphi1hour} and magenta vertical bar in Figure~\ref{fig:phitimeint}) have negative $\Delta\phi$, and vice versa (cyan arc and vertical bar). It should be noted that this definition of $\Delta\phi$ is consistent with that of \citet{Lario2013}, whereas \citet{Richardson2014} uses the opposite sign.

Figure~\ref{fig:phitimeint} shows several interesting features of the particle distribution and its evolution in time. The particles arrive rapidly, within the first 35 minutes, to a $\sim60^\circ$-wide range of longitudes, slightly shifted westwards from $\phi_{src}$. The first particles arrive at around $\Delta\phi\approx-15^\circ$ (see the black curve in Figure~\ref{fig:phitimeint}~(a) which shows the longitude of the peak intensity as a function of time). The highest intensity is seen at $\Delta\phi\approx-10^\circ$ an hour after the injection of the particles (red circle in Figure~\ref{fig:phitimeint}~(a)). At later times, as the intensities begin to decay, the particle population begins to broaden in heliolongitude, and after 48~hr, the particles cover over $200^\circ$ range of heliolongitudes, with the highest intensity at longitudes $\Delta\phi\approx0^\circ$ to $-5^\circ$. It should be noted that as our model does not include the convective electric field, the effect of corotation drift on the SEP time-intensity profiles reported in \citet{Hutchinson2023_Corot} is not reproduced in our results. We also note that the mean colatitude of the particles evolves from the initial $75^\circ$ to $85^\circ$ during the 48~hr simulation period (not shown), that is, the particles experience southwards gradient and curvature drifts \citep{Marsh2013,Dalla2013}.

In Figure~\ref{fig:phitimeint}~(b), we show the time-density profiles at three observers, one connected to the source at $\Delta\phi=0^\circ$ (black dash-dotted curve), one westward at $\Delta\phi=-30^\circ$ (magenta solid curve), and one eastward $\Delta\phi=30^\circ$ (blue dashed curve) from the field-line connected to the source. The onset phase is clearly asymmetric: the density starts to rise simultaneously at $\Delta\phi=0^\circ$ and $\Delta\phi=-30^\circ$, 4~minutes earlier than at $\Delta\phi=30^\circ$. Subsequently, the time-intensity profiles at the eastward and westward locations reach similar values, whereas at the the peak intensity at $\Delta\phi=0^\circ$ exceeds that of the east- and westward locations by an order of magnitude.

\begin{figure}
  \resizebox{\hsize}{!}{\includegraphics{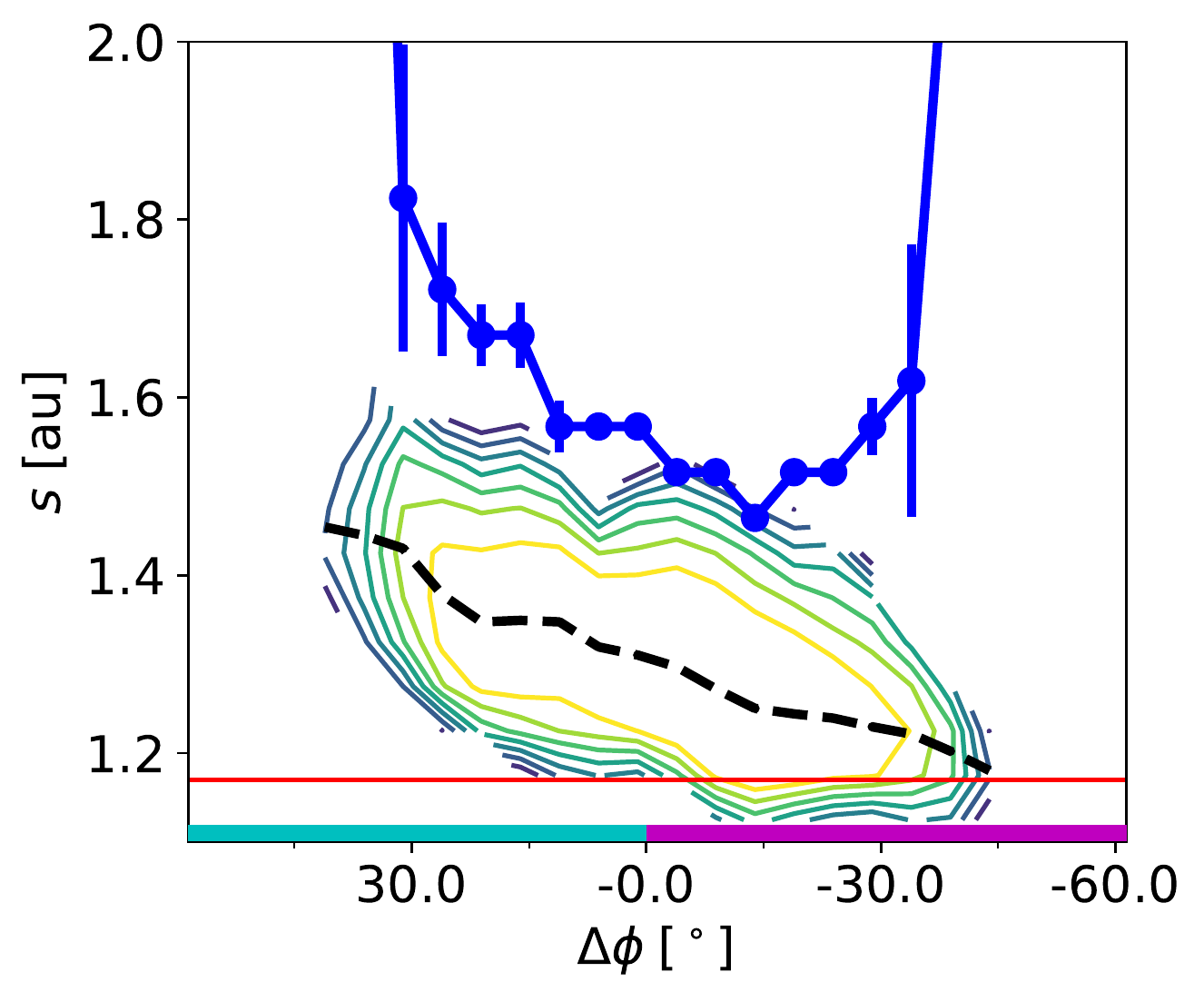}}
  \caption{Field line length density (contour lines, spaced at 0.5 orders of magnitude \citep{LaEa2022_ParkerTurb}) and path lengths of the first arriving 100-MeV protons (thick blue curve)) as a function of $\Delta\phi$. The arrival times, and hence the path lengths, of the protons is determined as the time when the intensity reaches 1\% of the peak intensity at 1~au. The black dashed curve shows the mean field line length, and the horizontal red line the nominal Parker spiral length. The magenta and cyan ranges at the bottom correspond to those in Figure~\ref{fig:rphi1hour}.}\label{fig:phi_onset}
\end{figure}

We investigate the onset time of the simulated SEP event at 1~au further in Figure~\ref{fig:phi_onset}, where we define the onset time as the first time the SEP intensity exceeds a threshold of 1\% of the global peak intensity of the simulated event, $n=1.5\;\mathrm{particles}/\left(\mathrm{s}\;^\circ\right)$ at $\phi=-52^\circ$ at 1~au. We further convert the onset time to an apparent path length (solid blue curve in Figure~\ref{fig:phi_onset}), for comparison with the field line lengths obtained in \citet{LaEa2022_ParkerTurb} for the same turbulence parameters as in this study (contour lines, with the dashed black line showing the mean field line length), and the nominal Parker spiral length (the horizontal red solid line). 

As can be seen in Figure~\ref{fig:phi_onset}, the apparent path length of the 100~MeV protons traces the general trend of the field line lengths in that within the range of $\Delta\phi=-15^\circ$ to $15^\circ$ the westward path lengths are considerably shorter than the eastward ones. However, the path lengths are 20\% longer than the mean field line length (black dashed curve in Figure~\ref{fig:phi_onset}), and 30-50\% longer than the nominal Parker spiral length. At more western observer locations, $\Delta\phi<-15^\circ$, the apparent SEP path length begins to increase, deviating from the trend of decreasing field line lengths. Similar lengthening of the apparent SEP path length, compared to the mean field line length, can be seen at eastern longitudes $\Delta\phi>15^\circ$.


\begin{figure}
  \resizebox{\hsize}{!}{\includegraphics{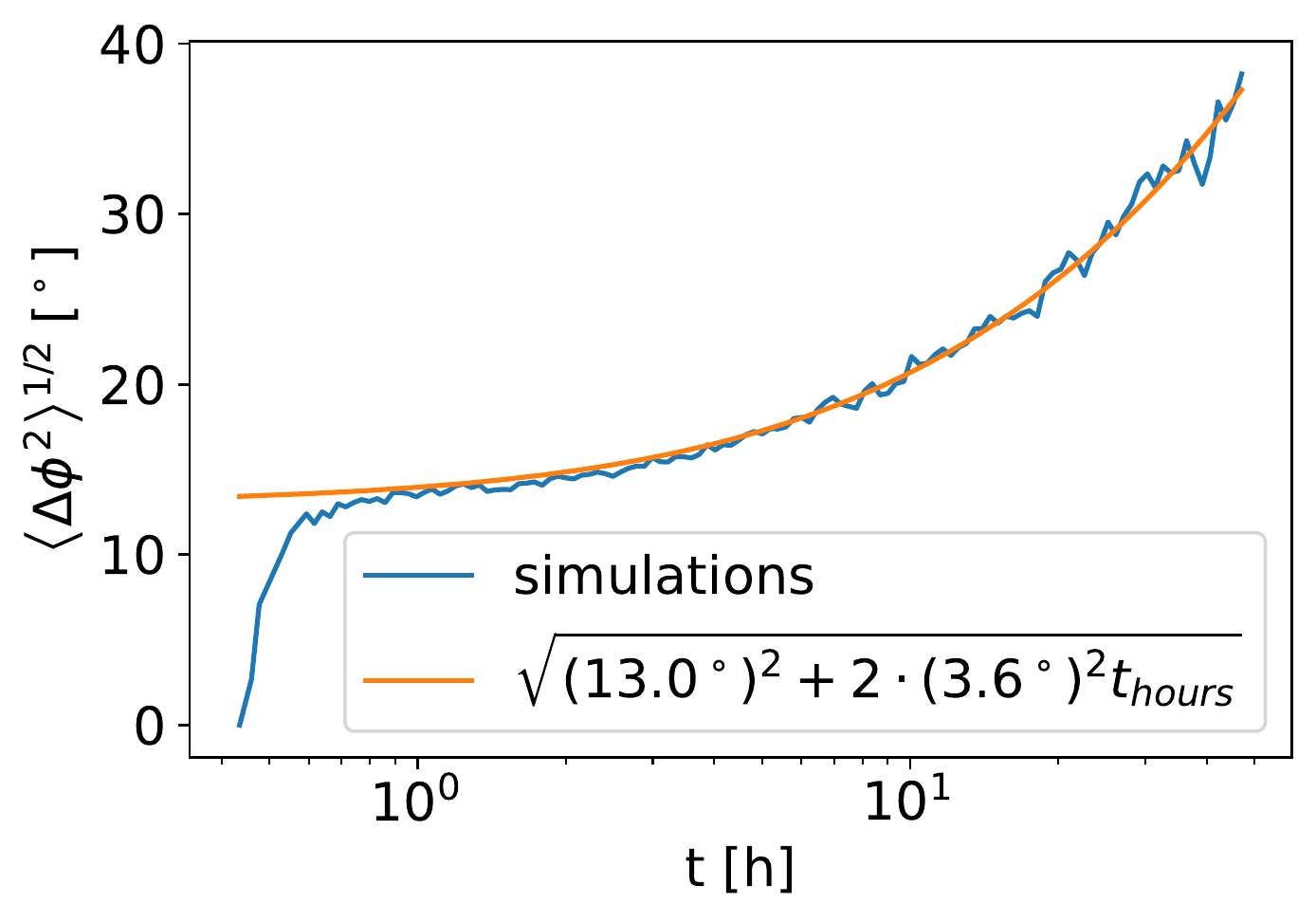}}
  \caption{Heliolongitudinal standard deviation of the 100-MeV protons as a function of time. The amber curve shows the functional form where the standard deviation is initially $13^\circ$, and then increases consistent with longitudinal diffusion with diffusion coefficient $\kappaang=(3.6^\circ)^2/\mathrm{hr}$.}\label{fig:stdev}
\end{figure}

After the rapid SEP onset at broad range of heliolongitudes, the longitudinal distribution of the SEPs broadens gradually, resulting in considerably longer onset delays at wider longitudes $\left|\Delta\phi\right|\gtrsim 40^\circ$ (see Figure~\ref{fig:phitimeint}). We trace the broadening of the distribution in Figure~\ref{fig:stdev}, where we have calculated the temporal evolution of the longitudinal standard deviation of particles at 1~au (blue curve). As can be seen, after the rapid increase of the standard deviation to $\sigma_\phi\approx 13^\circ$ within the first tens of minutes after the first particles arriving, the distribution begins to gradually increase. We fit the standard deviation with  
\begin{equation}\label{eq:stdev_time}
 \sigma_\phi^2(t)\equiv\left<\phi^2\right>_{r=1au}=\sigma_{FL}^2+2\kappaang t,    
\end{equation}
which describes the evolution of the variance of the diffusing particle population, with diffusion coefficient of $\kappaang$ and initial distribution being a Gaussian with standard deviation of $\sigma_{FL}$. A fit done by eye  gives $\sigma_{FL}=13^\circ$, which corresponds to the standard deviation of the meandering magnetic field line distribution obtained for our turbulence parameters in \citet{LaEa2022_ParkerTurb}, and particle longitudinal diffusion coefficient $\kappaang=(3.6^\circ)^2/\mathrm{hr}$ (orange curve). The close match between the two curves shows that the longitudinal evolution of the particle distribution can be described as diffusion of particles from an initial distribution that is determined by the random-walk of field lines as they traverse the distance between the Sun and 1~au.  It should be noted that the above analysis does not take into account the effect of longitudinal drifts which are minimised by injecting particles at latitude 0 degrees. However, as particles move away from this latitude some drift in longitude will take place \citep{Dalla2013}.

\section{Discussion and conclusions}\label{sec:discussion}

In this paper, we have presented the first Solar Energetic Particle simulations in heliospheric configuration where the large-scale Parker Spiral magnetic field is superposed with a predominantly 2D turbulent component that is two-dimensional with respect to the background magnetic field. The turbulence model was presented in detail in  \citet{LaEa2022_ParkerTurb}, where we showed that the turbulently meandering field lines were significantly longer than the Parker spiral, and longer at observer locations connected eastward from the source, as compared to westward observers (coloured contours in Figure~\ref{fig:phi_onset}).

The study presented in \citet{LaEa2022_ParkerTurb} analysed only the field line lengths, thus not accounting for the effects of the turbulence on SEP propagation along and across the meandering field lines. In this work, we analysed the propagation of 100~MeV protons using full-orbit test particle simulations, where the transport effects due to turbulence are naturally taken into account. 

We found that the protons have a rapid access to longitudinal range of 60$^\circ$ within first tens of minutes after the first arrival of the particles to 1~au. A wide range of longitudes accessed by SEPs rapidly has been reported by, e.g., \citet{Richardson2014} and \citet{Dresing2014}.

The apparent SEP path lengths derived from the simulations are 30-50\% longer than the Parker spiral, and 20\% longer than the mean length of the field lines \citep{LaEa2022_ParkerTurb} within $|\Delta\phi|<15^\circ$. For larger $|\Delta\phi|$ the difference between the SEP path and field line length increases further, exceeding 30\% increase at $\phi|\gtrsim 30^\circ$. The lengthening of the SEP travelled length with respect to the field line lengths is due to turbulence-induced scattering of the particles along the meandering field lines. The shortest SEP path length is at $\Delta\phi=-15^\circ$. The path lengths and thus the onset times increase monotonically either side of this heliolongitude, consistent with trend seen in observations \citep[e.g.][]{Richardson2014}.

Our simulations show that the location of maximum intensity is shifted westwards with respect to $\Delta\phi=0^\circ$. Fits to measured SEP peak intensities versus $\Delta\phi$ indicate similar westward location of the maximum of the gaussian at $\Delta\phi\sim -15^\circ$ \citep{Lario2013, Richardson2014}. 


The evolution of the width of the heliolongitudinal standard deviation has been reported in recent studies. \citet{Dresing2018} find that during the 26 December 2013 event the standard deviation of the longitudinal intensity distribution of 30-60~MeV protons increased from $35^\circ$ to $50-60^\circ$ in 23 hours. For the evolution of the longitudinal standard deviation that follows the form given by Equation~(\ref{eq:stdev_time}) with $\sigma_{FL}=35^\circ$, this corresponds to $\kappaang\sim (6^\circ)^2/\mathrm{hr}$. Similar values can be obtained from the analysis  of four SEP events by \citet{Kahler2023} whose linear change rate of $\sim 10^\circ/\mathrm{day}$ are equivalent of diffusive rates $\sim(5^\circ)^2/\mathrm{hr}$. These values are slightly higher than our results, however they are of the same order, indicating that turbulence-induced diffusion of SEPs across the mean field, as seen in our simulations, is a possible explanation for the observed evolution of the cross-field extent of SEPs during solar events.

It should be noted that the turbulence parameters determine the degree of field line meandering, and thus the initial extent of the SEP event ($\sim60^\circ$ in Figure~\ref{fig:phitimeint})  \citep[e.g.][]{LaEa2017meandstatistic}. Parameters such as those used in \citet{LaEa2016parkermeand} and \citet{Chhiber2021_randomwalk} likely result in a wider extent. 

We conclude that several features of observed SEP events can arise naturally from the turbulent nature of the interplanetary magnetic field. This does not preclude other suggested mechanisms, such as wide CME source \citep[e.g.][]{Reames1999} which has been suggested to be the reason for wide SEP events, as well as to shift the SEP peak intensity  \citep{Lario2013, Kahler2023}, and to temporally increase the heliolongitudinal extent of an SEP event \citep[e.g.][]{Kouloumvakos2016}. Further work, with a wide range of turbulence and particle source parameters, are needed to analyse the relative importance of different mechanisms in forming the observed SEP intensity evolution at different heliolongitudes.

\begin{acknowledgements}
TL and SD acknowledge support from the UK Science and Technology Facilities Council (STFC) through grants ST/R000425/1 and ST/V000934/1. CW and SD acknowledge support from NERC via the SWARM project,  part of the SWIMMR programme (grant NE/V002864/1). This work was performed using resources provided by the Cambridge Service for Data Driven Discovery (CSD3) operated by the University of Cambridge Research Computing Service (www.csd3.cam.ac.uk), provided by Dell EMC and Intel using Tier-2 funding from the Engineering and Physical Sciences Research Council (capital grant EP/P020259/1), and DiRAC funding from the Science and Technology Facilities Council (www.dirac.ac.uk). TL acknowledges support from the International Space Science Institute through funding of the International Team \#35 "Using Energetic Electron And Ion Observations To Investigate Solar Wind Structures And Infer Solar Wind Magnetic Field Configurations".
\end{acknowledgements}

\bibliographystyle{aasjournal}
\bibliography{ms}

\end{document}